\documentclass[aps,prb,citeautoscript,superscriptaddress,reprint,twocolumn]{revtex4}

\usepackage{graphicx}
\usepackage{amsfonts,amsmath,amssymb,amsthm}
\usepackage{epstopdf}
\usepackage{upgreek,xspace}
\usepackage{chngcntr}
\usepackage[colorlinks,allcolors=blue]{hyperref}
\usepackage[version=3]{mhchem} 

\newcommand{\be}{\begin{equation}}
\newcommand{\ee}{\end{equation}}

\newcommand{\rfive}{$\sqrt{5}\times\sqrt{5}$}
\newcommand{\rfiveo}{$\sqrt{5}\times\sqrt{5}\times1$}
\newcommand{\rfived}{$\sqrt{5}\times\sqrt{5}\times2$}

\begin{document}

\title{Supercell formation in epitaxial rare-earth ditelluride thin films}

\author{Adrian~Llanos}
\affiliation{Department of Applied Physics and Materials Science, California Institute of Technology, Pasadena, California 91125, USA.}
\affiliation{Institute for Quantum Information and Matter, California Institute of Technology, Pasadena, California 91125, USA.}

\author{Salva~Salmani-Rezaie}
\affiliation{School of Applied and Engineering Physics, Cornell University, Ithaca, New York 14853, USA}
\affiliation{Kavli Institute at Cornell for Nanoscale Science, Cornell University, Ithaca, New York 14853, USA.}

\author{Jinwoong~Kim}
\affiliation{Department of Physics and W. M. Keck Computational Materials Theory Center, California State University, Northridge, Northridge,
California 91330, USA.}

\author{Nicholas~Kioussis}
\affiliation{Department of Physics and W. M. Keck Computational Materials Theory Center, California State University, Northridge, Northridge,
California 91330, USA.}

\author{David~A.~Muller}
\affiliation{School of Applied and Engineering Physics, Cornell University, Ithaca, New York 14853, USA}
\affiliation{Kavli Institute at Cornell for Nanoscale Science, Cornell University, Ithaca, New York 14853, USA.}

\author{Joseph~Falson}
\email{falson@caltech.edu}
\affiliation{Department of Applied Physics and Materials Science, California Institute of Technology, Pasadena, California 91125, USA.}
\affiliation{Institute for Quantum Information and Matter, California Institute of Technology, Pasadena, California 91125, USA.}

\begin{abstract}
Square net tellurides host an array of electronic ground states and commonly exhibit charge-density-wave ordering. Here we report the epitaxy of DyTe$_{2-\delta}$ on atomically flat MgO (001) using molecular beam epitaxy. The films are single phase and highly oriented as evidenced by transmission electron microscopy and X-ray diffraction measurements. Epitaxial strain is evident in films and is relieved as the thickness increases up to a value of approximately 20 unit cells. Diffraction features associated with a supercell in the films are resolved which is coupled with Te-deficiency. First principles calculations attribute the formation of this defect lattice to nesting conditions in the Fermi surface, which produce a periodic occupancy of the conducting Te square-net, and opens a band gap at the chemical potential. This work establishes the groundwork for exploring the role of strain in tuning electronic and structural phases of epitaxial square-net tellurides and related compounds.
\end{abstract}


\flushbottom
\maketitle

\section*{Introduction}
Layered compounds sharing the square-net structural motif support a myriad of electronic phases including superconductivity\cite{haoFeSe}, topological protected modes,\cite{klemenz:2019} magnetism \cite{Sankar2019,lei:2021} , and charge density waves (CDW)\cite{Ru:2008chemicalpressure,shin:2005,Lei:2019}.Among this class of materials, rare-earth tellurides (LnTe$_x$, where Ln is a lanthanide and $x$ = 2, 3 most commonly), have attracted attention due to the interplay between itinerant carriers\cite{lei:2020highmobility} and magnetism, superconductivity and CDW-order.\cite{yumigetareview:2021} Sandwiched between corrugated quasi-ionic Ln-Te layers, the electronic structure of the covalently bonded two-dimensional square Te sheet can conceptualized by a partially filled p-orbital manifold which obtains extra charge from an imbalance in valence states within the corrugated layer \cite{Tremel:1987,hoffmann:1988}. Modifying Ln acts to exert chemical pressure and influences both the electronic properties of the material. For example, in the case of the rare-earth tritellurides, chemical pressure acts to modify the CDW wavevector $q$, the CDW-ordering temperature, and the orthorhombicity of the crystal, which ultimately reveals two orthogonal CDW with dissimilar ordering temperatures.\cite{Ru:2008chemicalpressure,yumigetareview:2021} Ultra-fast optical probes have demonstrated transient expression of competing CDW orders, suggesting a complex energy landscape governs their stability.\cite{kogar2020light} Studies on the effect of hydrostatic pressure have revealed that CDW order can be further suppressed and superconductivity being promoted, all in the presence of rare-earth-derived (antiferromagnetic) ordering.\cite{yumigetareview:2021}.

\begin{figure}[ht]
	\includegraphics[width=82mm]{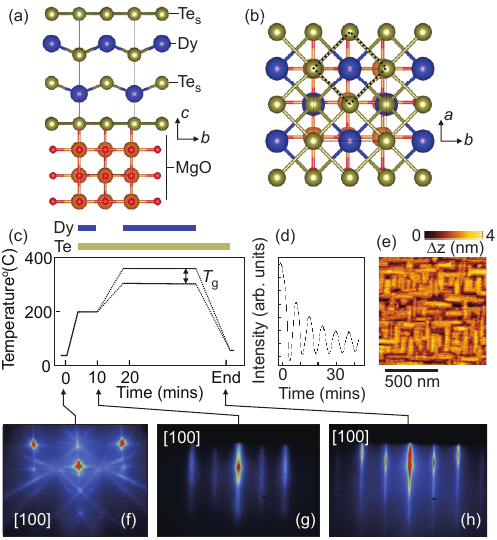}
	\caption{Crystal structure of DyTe$_2$ on MgO plotted along the (a) $a$ axis and (b) $c$ axis directions. (c) Schematic of film growth recipe indicating substrate temperature as a function of time. Color bars at the top of the panel indicate when the Dy and Te effusion cells are open. (d) RHEED oscillations from the specular reflection of films as a function of growth time. (e) AFM image of a $T_\mathrm{g}=330^\circ$C film of 20 u.c. RHEED of (f) thermally prepared substrate, (g) after buffer layer deposition, and (h) at the conclusion of growth, all recorded along the [001] direction of the MgO substrate.}
	\label{Fig1}
\end{figure}

The related LnTe$_2$ display similar behaviors to the tritellurides, but differ in key ways due to their electron filling of 7e-/Te in the square planar layers \cite{Tremel:1987, kikuchi:1998}. In contrast to the tritellurides, which form stoichiometric crystals, the ditellurides have a substantial width of formation and tend to form Te-deficient crystal structures with a range of stoichimetries and resulting superstructure modulations \cite{nihara:1972,Park_SmTe2,shin:2005,poddig:2018}. These modulations are often commensurate with the parent structure and result from an ordering of Te vacancies within the Te sheet. However, incommensurate modulations have also been observed in LaTe$_{2-\delta}$ and CeTe$_{2-\delta}$ \cite{shin:2005}. Electrical transport ranging from metallic \cite{stowe:2000} to semiconducting \cite{Jung:1997,Kwon:2000} for similar nominal compositions and photoemission studies reporting both pseudogap \cite{shin:2005,chung:1999} and fully-gapped \cite{garcia:2007} Fermi surface structures further demonstrate the complexity of the ditelluride system. Superconductivity has also been reported in CeTe$_2$ at 0.2~GPa\cite{Jung:superconCeTe2} and in Sb-doped LaTe$_2$ at approximately 3~GPa\cite{chen2022}. The subtle dependence on defect stoichimetry and applied pressure motivate an investigation into the role that epitaxial strain and dimensionality play in the stabilization of lattice distortions and the resulting groundstate of the electron-phonon system. 

Here we report the epitaxial growth of rare-earth ditelluride DyTe$_{2-\delta}$ on MgO (001) substrates. While little is known about DyTe$_{2-\delta}$ in the bulk, the closely related structures of LaTe$_{2-\delta}$ and CeTe$_{2-\delta}$ \cite{shin:2005} are are reported to either have tetragonal or orthorhomic\cite{dimasi:1996} symmetry belonging to the $P4/nmm$ and $Pnma$ space groups, respectively. We have found in all experiments that the ditelluride phase is stabilized preferentially over the tritelluride phase despite synthesis being performed in Te-rich conditions. We attribute this to the weak van-der Waals interactions between stacked Te-sheets on the surface of samples during layer-by-layer growth. While growth of the tritelluride phase has been reported in the ultra-thin limit \cite{Xu_2021} ($<$ 1 unit cell (u.c.)), this appears to not persist into multilayer films, even as thin as 3 u.c. 
Our results identify that epitaxial strain is evident in films and is dependent on the number of quintuple layers (QL) deposited, being progressively relieved as the films are made thicker. The films are tetragonal within the detection limit of the X-ray diffraction apparatus. We additionally have resolved a commensurate superlattice in a subset of films with wave vectors $\vec{q_1} = \frac{2}{5}\vec{a}^* + \frac{1}{5}\vec{b}^*$\ ,\ $\vec{q_2} = -\frac{1}{5}\vec{a}^* + \frac{2}{5}\vec{b}^*$\ text{and}\ $\vec{q_3} = \frac{1}{2}\vec{c}^*$ where $\vec{a}^* ,\vec{b}^* \text{and}\ \vec{c}^*$ are the reciprocal lattice vectors of the undistorted structure. In real space, this distortion can be thought of as ($\sqrt{5}\times\sqrt{5}$)R26.6$^\circ$~$\times$~2 superlattice. In the case of Ref.\citenum{poddig:2018}, the distortion was associated with a reduction in symmetry of the compound from the $P4/nmm$ to $P4/n$ space group. The superlattice emerges due to a periodic Te-defect lattice, which also acts to open a gap in the electronic spectrum and induce semiconducting transport behaviour. First principles calculations point towards nesting conditions of the Fermi surface at a $q$-vector that corresponds to the \rfive~condition, suggesting that the formation of the defect lattice results from a similar driving force to the conventional picture of CDW formation where sections of the Fermi surface are gapped out by the formation of supercells with periodicity corresponding to the nesting condition\cite{lee:1996}. It is therefore a thermodynamically favored process rather than an outcome of imperfect growth conditions. The calculations further reveal that a number of competing atomic arrangements may be present and dependent on the degree of strain in the crystal.

\section*{Experimental Methods}
Substrates are prepared by annealing in ultra-high vacuum (base pressure $\approx 10^{-10}$ Torr) using a home-built CO$_2$ laser heating apparatus, inspired by Ref.\citenum{braun2020}. We used MgO (001) (Crystec) substrates for growth without back coating of an absorbing layer or additional cleaning steps. The combination of MgO ($a$=0.421~nm) and DyTe$_2$ ($a$=0.428~nm) was selected due to the relatively good epitaxial relationship at approximately -1.6\% mismatch, as shown in Fig.\ref{Fig1}(a) and (b) where we highlight the DyTe$_2$ unit cell and its constituent atoms; sheet tellurium (Te$_\mathrm{s}$), the corrugated layer tellurium (Te$_\mathrm{c}$), and the rare-earth ion. Being hygroscopic, the MgO surface without annealing is of poor quality, but upon heating to in excess of 1000~$^\circ$C we can generate step-terraced structures within a few minutes\cite{braun2020}. The substrate is then transferred to a dedicated telluride substrate holder in a glove box, and then reloaded into the vacuum cluster and transferred to the molecular beam epitaxy growth chamber. The reflected high energy electron diffraction (RHEED) signal from the surface remains qualitatively the identical upon arrival into the growth chamber. Substrate heating in the growth chamber is provided by a conventional resistive SiC coil, and the temperature reported is read from a thermocouple between the substrate and heating coil. Te was supplied using a thermal cracker cell with tank temperatures of approximately 450~$^\circ$C and cracking zone temperature of 1000$^\circ$C. Dysprosium was evaporated from a conventional effusion cell at a temperature of 1075~$^\circ$C. We have found that both thermal cracking of Te and thermal preparation of substrates play a role in improving sample quality. Due to the air sensitivity of the films, we cap each film before extracting them from the vacuum chamber with either amorphous Ge or partially crystalline Te. This is performed \textit{in-situ} at $T\approx$90~K using a liquid nitrogen cooling stage. X-ray diffraction data was obtained using a Rigaku Smartlab diffractometer using a 2-bound Ge (220) monochromator. High resolution diffraction employed copper $K\alpha1$ radiation. Diffraction in the (hk0) plane (i.e. the $\chi$-circle) was accessed using an in-plane diffraction arm in an incidence angle $\omega ~ 0.35^{\circ}$. A $0.5^{\circ}$ parallel slit collimator and $0.5^{\circ}$ parallel slit analyzer were used to reduce in-plane beam divergence. 

The cross-sectional Scanning Transmission Electron microscopy (STEM) samples are prepared using a standard lift-out process on Helios G4 UX DualBeam focused ion beam system. We utilize the Spectra 300 X-CFEG operating at 200 kV with a semi-convergence angle of 30mrad and a High-Angle Annular Dark-Field (HAADF) detector with an angular range of 60-200 mrad to collect HAADF-STEM images. The HAADF-STEM images are acquired as a series of 25 images (250 ns per frame) and subsequently averaged to produce images with a high signal-to-noise ratio. STEM- Energy dispersive X-ray spectroscopy (EDX) data is acquired using a steradian Dual-X EDX detector, and the resulting spectrum is denoised through the application of the principal component analysis (PCA).
We use an electron microscope pixel array detector (EMPAD) to collect Convergent beam electron diffraction (CBED) patterns at a convergence angle of 1.2 mrad and dwell time of 1 ms. For the CBED simulation, we use the Py4DSTEM package based on the Bloch wave method.\cite{Blochmethod}

\section*{Calculation Methods}

The density functional theory calculations are performed using the Vienna $ab$ $intio$ simulation package\cite{Kresse96a,Kresse96b} with the projector augmented wave (PAW) method.\cite{Blochl94,KressePAW} The Dy-4$f$ states are treated as core. All structures are optimized using the PBEsol exchange correlation functional\cite{PBEsol} and the Bloch states are wannierized\cite{Mostofi,Marzari2012} with the modified Becke-Johnson (mBJ) potential\cite{mBJ1,mBJ2} which provides accurate band gaps, effective masses and frontier-band ordering. The Fermi surfaces and nesting functions (supplementary information (SI)) are obtained using the tetrahedron method with the Wannier Hamiltonian.

\section*{Results and Discussion}

The growth procedure is illustrated in Fig\ref{Fig1}(c). We begin with a low temperature ($T$=200$^\circ$C) buffer layer on the annealed MgO surface. By observing intensity oscillations of the RHEED pattern, precise thickness control can be realized allowing for the buffer layer growth to be stopped after a single QL. Attempts to grow films without a buffer layer resulted in poorly crystalline films, as observed by weak, diffuse features in RHEED. Growth of the remainder of the film takes place after annealing the buffer under Te flux to a final growth temperature. Growth rate is set by the flux of Dy atoms and was set at ~1.2 A/min for all growths under Te over pressure. Te/Dy flux ratio were typically in the range of 10 to 20, corresponding to beam flux pressures of 7$\times 10^{-9}$~mbar for Dy and 1.4$\times 10^{-7}$~mbar for Te. We have deposited films within a range of growth temperatures ($T_\mathrm{g}$), as shown in \ref{Fig1}(c). For films deposited at low temperature ($T_g<$300$^\circ$C), rough island growth appears whereas those grown at elevated ($T_g>$~350$^\circ$C) overall suffer from poor crystallinity and rough surfaces, likely due to the low sticking coefficient of Te. RHEED oscillations during growth for moderate growth temperatures ($T_g=$~315$^\circ$C, Fig.\ref{Fig1}(d)) indicate that the film grows layer-by-layer with oscillations corresponding to the growth of a single unit cell. These oscillations were observed to decay after approximately 10 oscillations and were completely damped by the completion of a 20uc thick film. To estimate the growth rate, RHEED oscillations were observed for the first several layers. The substrates are subsequently rotated at 5 RPM to enhance sample uniformity. Fig.\ref{Fig1}(e) shows an atomic force microscopy image (AFM) of the final surface of a 20 u.c. film, showing surface roughness on the order of 2~3 u.c. RHEED images are shown of the pristine substrate (Fig.\ref{Fig1}(f)), after buffer layer (g) and post-growth (h), the latter of which shows sharp, streaky patterns with prominent Kikuchi bands. 

\begin{figure}[ht]
	\includegraphics[width=85mm]{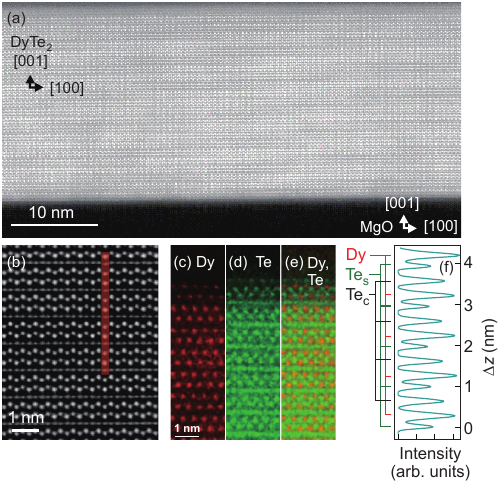}
	\caption{HAADF imaging of the film from a wide angle (a) and at the interface with MgO (b). (c-d) EDX mapping of the film showing occupancy of Dy and Te. (f) Intensity of atoms in a vertical line cut of the HAADF data indicated as the red bar in panel (b), showing the oscillatory intensity as a function of distance. The Dy, Te$_s$ and Te$_c$ sites are noted.}
	\label{Fig2}
\end{figure}

Transmission electron microscopy analysis is presented in Fig.\ref{Fig2} and shows stacked layers of planar Te sheets and corrugated Dy-Te interlayers. The stacking sequence confirms the presence of the DyTe$_2$ phase and not DyTe$_3$. We attribute this to the low sticking coefficient of the Te-sheets, which evidently cannot be stacked in pairs (as is the case in the tritellurides) even with a Te/Dy flux ratio as high as 40. The film is phase pure with a sharp interface formed between the MgO and Te-square net sheet (Fig.\ref{Fig2}(b)). The square net tellurium sheet forms the first wetting layer on the substrate. Rotated domains are evident in some areas of the film, but are suppressed after a few monolayers of growth. The high-angle annular dark-field imaging (HAADF) (Fig.\ref{Fig2}(a,b)) and energy-dispersive X-ray spectroscopy elemental map (Fig.\ref{Fig2}(c-e)) shows nominal (P4/nmm) DyTe$_2$ structure. The Dy/Te ratio with absorption and ionization cross-section correction for the foil with a thickness of 30 nm is 0.497. This ratio, at its best, has a 10-20\% error because of strong channeling.\cite{lugg:2015,macarthur:2017} When the beam is along the zone axis, the intensity of x-ray spectra is not a linear contribution of each element. The experimental CBED pattern (see SI) matches the simulated CBED pattern of the ($P4/nmm$) point group when recorded along the [010] direction. The high resolution images also capture dissimilar intensity of Te sites along a line profile in the DyTe$_2$ [001] direction, as plotted in Fig.\ref{Fig2}(f). The intensity of the Te columns at the Te sheet (Te$_s$) and the Te columns at the corrugated layer (Te$_c$) is different. We attribute this to the phase space for vibrations of the atoms in the square sheet versus those in the ionic Dy-Te layer; the higher thermal vibration in the Te sheets is an indication that the atoms are loosely bound. This observation is in agreement with our experience of being unable to stabilize the tritelluride phase as a thin-film due to the poor sticking coefficient of Te atoms in the square net layer.

\begin{figure}[ht]
	\includegraphics[width=85mm]{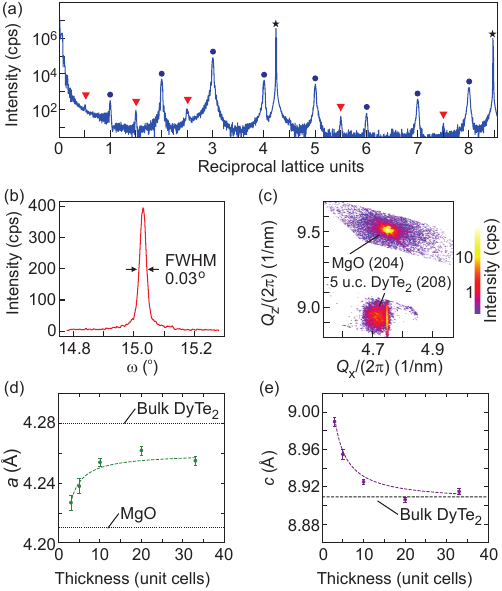}
	\caption{(a) Out-of-plane $\theta$-2$\theta$ X-ray diffraction of a 20 u.c. DyTe$_{2-\delta}$ film on MgO (001), plotted in reciprocal lattice units of DyTe$_{2-\delta}$. Peaks corresponding to the substrate are indicated by a star, with fundamental peaks of the film labelled by a circle and superlattice peaks by a triangle. (b) Rocking curve analysis of the DyTe$_{2-\delta}$ (003) diffraction peak. (c) Out-of-plane asymmetric scan of the film with diffraction from the MgO (204) and DyTe$_{2-\delta}$ (208) planes plotted as a function of momentum transfer in the $Q_x$--$Q_z$ plane. (d) In-plane $a$ and (e) out-of-plane $c$ lattice constants of films as a function of thickness. Dashed lines are fits intended as guides for the eye.}
	\label{Fig3}
\end{figure}

The structural properties of grown films were further investigated using X-ray diffraction. Figure \ref{Fig3}(a) shows a $\theta$-$2\/\theta$ scan, plotted in reciprocal lattice units, showing highly $c$-axis oriented films with intense layer peaks accompanied by prominent Laue fringes indicating a high quality film with a sharp interface with the underlying substrate \cite{pietsch}. An $\omega$ rocking curve analysis of the DyTe$_2$ (003) peak produces a full-width at half maximum (FWHM) of 0.03$^\circ$ indicating low out-of-plane mosaicity. $\phi$ scan data (not shown) around the (110) peaks taken in grazing incidence geometry reveals the four-fold symmetry of the layer and further confirms the $[100]_\mathrm{DyTe_2} || [100]_\mathrm{MgO}$ epitaxial relationship observed with RHEED.

\begin{figure}[ht]
	\includegraphics[width=75mm]{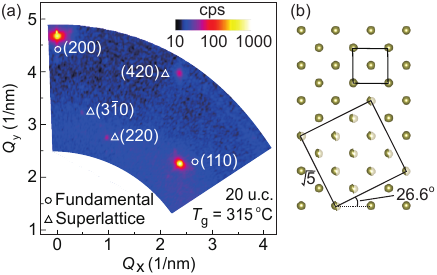}
	\caption{(a) In-plane (hk0) diffraction of a 20 u.c. film showing diffraction peaks identified as fundamental (circle) and superlattice (triangle). (b) Schematic of the Te square net layer showing the original square u.c. (upper square) and the distorted \rfive~R26.6$^\circ$ superlattice (lower rotated square) of partially occupied Te-sites, indicated by half-transparent atoms.}
	\label{Fig4}
\end{figure}

One goal of this work is to investigate the presence of epitaxial strain and its dependence on film thickness and growth parameters. To this end, we employ high resolution reciprocal space mapping in the $Q_x$--$Q_z$ plane for films with thickness between 3 and 35 u.c. A characteristic data set for a 5 unit cell film is plotted in Fig.\ref{Fig3}(c), showing the MgO (204) and DyTe$_{2-\delta}$ (208) diffraction spots. For the 3 unit cell film, intensity in the $Q_x$--$Q_z$ plane was too weak to fit a peak. In this case, measurements of the $Q_x$--$Q_y$ RSM were performed about the $(hh0)$ for $odd\ h$ layer diffraction peaks in grazing incidence geometry. These reflections were chosen to ensure that the MgO substrate would not contribute to the measurement. The $c$-axis lattice constants were determined from the $\theta-2\theta$ measurements. In Fig\ref{Fig3}(d) and (e) we plot the evolution of the measured $a$ and $c$ lattice constants for films grown under a Te/Dy flux ratio of 20 and a growth temperature $T_g$=310$^\circ$C. There is a continuous evolution of in-plane compressive strain from -$1.23\%$ to $0.5\%$ relative to reported bulk values for DyTe$_{1.8}$. As will be described in the subsequent section, the presence of the observed superlattice implies that the stoichiometry is closest to DyTe$_{1.8}$ and the epitaxial strain estimates were referenced to these values \cite{poddig:2018}. The relaxed value of the in-plane lattice constant was found to be 4.26$\AA$ which. As expected for films showing in-plane compressive strain, the $c$-axis lattice constant was observed to evolve from $+1.1\%$ at the lowest thickness to $+0.1\%$ by 20nm. Films grown at higher temperatures ($T_g>$ 330$^\circ$) showed no evolution of strain with thickness. 

It is known that in the case of La and Ce the ditelluride structure is stable over a range of stoichiometries in the bulk, typically within the range of ${x}=1.8\sim2.0$. The effects of such off-stoichiometries in the case of LaTe$_{1.8}$ leads to a gap opening at the chemical potential.\cite{poddig:2018} The off-stoichiometries are known to produce a variety of structural modulations depending on Te deficiency. The most commonly observed modulation, for a Te deficiency of 20\%, was a (\rfive$\times 2$) superlattice of the $LnTe_2$ aristotype. The u.c. of this ordered defect modulation is oriented parallel to the [210] direction the higher symmetry structure. Such a defect modulation is reminiscent of that observed in the insulating phase of Fe-deficient FeSe \cite{liu:2020}  and in certain oxygen deficient surface reconstructions of SrTiO$_3$ \cite{newell:2006}.

\begin{figure}[ht]
	\includegraphics[width=85mm]{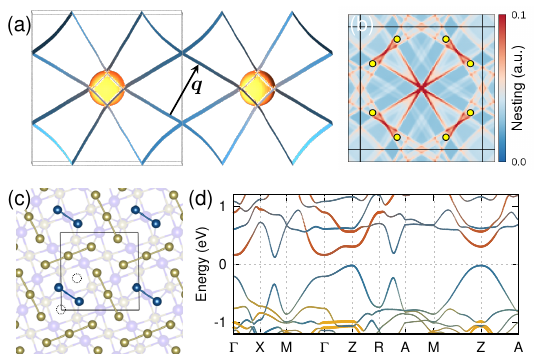}
	\caption{Density functional theory calculations of DyTe$_{2-\delta}$: (a) Fermi surface of pristine DyTe$_2$ at a chemical potential corresponding to $\delta=0.2$ Te deficiency, where \textbf{q} is the nesting vector. (b) In-plane nesting function, $N_{\gamma}(q_x,q_y)$, (see SI) in the first Brillouin zone (solid lines), where the eight yellow dots denote the nesting vector \textbf{q}, corresponding to the $\sqrt{5} \times \sqrt{5}$ cell modulation. (c) Te bonding network for $\delta=0.2$ where the Te dimers are colored as blue. (d) Band structure along the symmetry directions of the $\sqrt{5} \times \sqrt{5}$ cell shown in (c) revealing the emergence of the band gap. The orbital characters of the Dy-$d$, Te$_c$-$p$, and Te$_s$-$p$ derived states in (a) and (d) are denoted by orange, yellow, and blue, respectively.}
	\label{Fig5}
\end{figure}

Using grazing incidence X-ray diffraction we are able to analyze the structural modulations present in DyTe$_{2-\delta}$ thin films, as shown in Fig.\ref{Fig4}(a) where we plot diffraction of a 20 unit cell film in the (hk0) plane. The enhanced intensity in grazing incidence geometry for thin films enables use to resolve the presence of the $\sqrt{5}\times\sqrt{5}$ modulation of the a-b plane, indexed by the diffraction spots identified by triangles in the figure, which are absent in the undistorted structure. Diffraction indices given by triangles are with respect to the full supercell whereas those identified by circles are referenced to the undistorted structure. The presence of the (310) reflection in this region of reciprocal space indicates the presence of a rotated domain with modulation wavevector $\vec{q_1} = \frac{1}{5}\vec{a}^* + \frac{2}{5}\vec{b}^*$\ ,\ $\vec{q_2} = \frac{2}{5}\vec{a}^* - \frac{1}{5}\vec{b}^*$\ text{and}\ $\vec{q_3} = \frac{1}{2}\vec{c}^*$. As previously discussed, $\theta-2\theta$ diffraction measurements revealed additional features at 0.5 reciprocal lattice units, which correspond to a doubling of the unit cell along the $c$-axis. Additionally in RHEED, we are able to resolve weak streaks associated with an in-plane modulation (see Fig.\ref{FigOffaxis}). Taken all together, these diffraction features provide strong evidence for the ($\sqrt{5}\times\sqrt{5}$)~R26.6$^\circ$~$\times$~2 reconstruction being stabilized in the films. The presence of the defect lattice would imply an off-stoichiometry in the range of 10$\sim$20\%, which is consistent with the EDX-obtained chemical ratios within the uncertainty of the measurement. All films with thickness greater than 10nm showed features consistent with the 2-fold modulation of the $c$-axis. However, the comparably weaker diffracted intensity from films in the thin-limit combined with the intrinsically weak intensity from superlattice diffraction spots prevented the observation of the superlattice below 20 u.c. with our lab-based diffractometer. 

To further understand the underlying origin of the experimentally observed supercell, we have performed first principles calculations with and without Te deficiencies. 
Questions of interest include whether nesting within the Fermi surface promotes the formation of the supercell, the atomic arrangement(s) within the supercell, and the culminating band structure of the extended compound. 
Figure \ref{Fig5} presents the main results of our calculations, and an extended set of data is available in the SI. 
The Fermi surface of the unmodulated pristine DyTe$_{2-\delta}$ structure at the chemical potential corresponding to $\delta=0.2$ Te deficiency is presented in Fig.\ref{Fig5}(a) showing the Te$_s$ $p$- and the Dy $d$-derived bands in blue and orange, respectively. 
The planar shapes of the Te$_s$ derived Fermi surface confirm their 2D nature. 
The presence of bands at the Fermi energy would be expected to produce metallic transport in films. 
However, we observe semiconducting transport with an activation gap on the order of 300~meV (see SI). 
The gapping of the Fermi surface can be achieved through a number of means; in other square net compounds where CDW ground states have been identified, the Fermi surface hosts a large number of nesting vectors $q$ in momentum space, which produces enhancements in the Lindhard response function $\chi(\mathbf{q})$ when $q\approx2k_\mathrm{F}$. The induced charge is a product of 
the response function and the time-independent potential of the crystal. 
Thus, a highly nested Fermi surface, as is the case in the quasi-2D compound studied here, may drive the formation of CDW ground states.\cite{CDW_PNAS} 
The calculated nesting function, $N({\bf q})$, (see SI) of the Fermi surface in Fig.\ref{Fig5}(a) is displayed in Fig.\ref{Fig5}(b), where red
colors indicate wavevectors with a strong nesting condition at the Fermi energy. 
The yellow dots correspond to the $\mathbf{q}$-vector which would produce a \rfive~modulation, 
and they are seen to closely coincide with the nesting conditions of the Fermi surface. Therefore, we hypothesize that the \rfive~modulation is promoted in the structure to minimize energy. 
However, in contrast to purely electronic CDW-order, it appears likely that the crystal preferentially forms a defect lattice. Removal of a Te from the square net sheet effectively contributes extra charge to the layer, which in turn may be localized on the periodic defect site. The question remains what the defect lattice looks like locally and how it influences the band structure. 
Consequently, we have carried out calculations of the \rfive~supercell with 
Te$_s$ mono- and di-vacancies corresponding to $\delta=0.1$ and $0.2$, respectively. 
We find that the A-C divacancy configuration, consisting of Te$_s$ dimers and trimers, shown in Fig. \ref{Fig5}(c) has the lowest formation energy among the five considered structures and is in agreement with previous X-ray diffraction measurements on SmTe$_{1.8}$.\cite{Ijjaali_SmTe:2006} This configuration is favored in the calculations when the $a$-axis lattice parameter is compressed, which is a trend that is resolved in diffraction in Fig.\ref{Fig3}(d). The band structure of the A-C configuration along symmetry directions, shown in Fig. 5(d), demonstrates the emergence of band gap  (indirect gap of 143 meV and direct gap of 326 meV). We also find a band gap for the mono-vacancy and A-B di-vacancy configurations, while the A-D di-vacancy configuration with a higher formation energy is metallic.
Therefore, we speculate that these vacancy configurations, especially A-C, are stabilized and are responsible for the semiconducting transport behavior.
Additional structure modulation along the $c$ direction is also reproduced under the assumption of the dominant $\delta=0.2$ A-C configuration. The calculated total energies of the \rfived~structure (Table~\ref{Tab_c_mod}) show that the A-C divacancy appears at a laterally displaced position $(0.5, 0.5)$, relative to the divacancy on the adjacent Te square net layer. This lateral displacement lowers the total energy by approximately 10 meV/Dy compared to the \rfiveo~structure, where the A-C dimers align as a column.

\section*{Conclusion}
In summary, we have grown epitaxial DyTe$_{2-\delta}$ on atomically flat MgO surfaces which induces compressive strain in the films. The in-plane lattice constant of the films increases as the thickness is increased, indicating a relieving of strain. The high structural quality of films enables detection of a \rfive$\times2$~R26.6$^\circ$ superlattice in diffraction. First principles calculations point towards nesting in the Fermi surface as a possible mechanism for promoting the supercell modulation, which in turn produces a band gap and semiconducting transport. This work sets the stage for studying the role epitaxial strain has on the broader class of LnTe$_{2-\delta}$ and related square net compounds, and ultimately their heterointerfaces.



\subsection*{Acknowledgements}
We thank Kaveh Pezeshki for help in the early stages of this project and appreciate discussions with Ian Fisher, Leslie Schoop and Anshul Kogar. J.F acknowledges funding provided by the the Air Force Office of Scientific Research (Contract number FA9550-22-1-0463), and partial funding provided by the Gordon and Betty Moore Foundation’s EPiQS Initiative (Grant number GBMF10638), and the Institute for Quantum Information and Matter, an NSF Physics Frontiers Center (NSF Grant PHY-1733907). N.K and J.F. acknowledge support from the NSF-PREP CSUN/Caltech-IQIM Partnership (Grant number 2216774) and NSF-PREM (Grant number DMR-1828019). This work made use of a Helios FIB supported by NSF (Grant No. DMR-1539918) and the Cornell Center for Materials Research (CCMR) Shared Facilities, which are supported through the NSF MRSEC Program (Grant No. DMR-1719875). The Thermo Fisher Spectra 300 X-CFEG was acquired with support from PARADIM (NSF MIP DMR-2039380) and Cornell University. S.S-R. acknowledges the support of the Kavli and PARADIM fellowship. 
\clearpage

\section*{}
\bibliography{bib}

\clearpage

\renewcommand{\thefigure}{S\arabic{figure}}

\setcounter{figure}{0}

\subsection*{Supplementary Data}

\begin{figure}[h]
	\includegraphics[width=55mm]{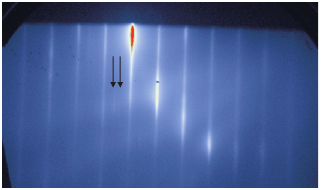}
	\caption{Off-axis RHEED image with saturated brightness showing weak additional streaks which we tentatively associated with the \rfive~modulation.}
	\label{FigOffaxis}
\end{figure}

\begin{figure}[h]
	\includegraphics[width=65mm]{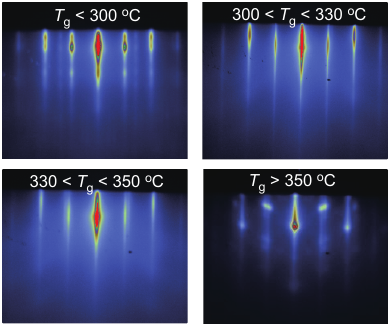}
	\caption{RHEED images taken at the end of growths performed in the indicated temperature windows.}
	\label{FigRHEED}
\end{figure}

\begin{figure}[h]
	\includegraphics[width=35mm]{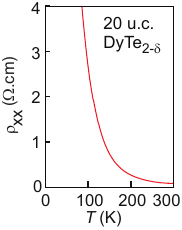}
	\caption{Temperature dependent transport of a 20~u.c. film showing semiconducting transport.}
	\label{FigSemicond}
\end{figure}

\begin{figure}[h]
	\includegraphics[width=85mm]{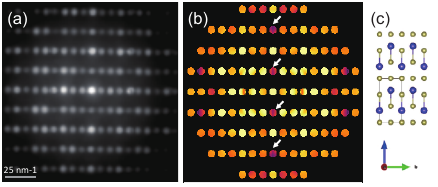}
	\caption{(a) CBED pattern which matches the (b) simulated CBED pattern of the (p4/nmm) point group.
\{0k0\} (k=odd) reflections, marked by the arrow, are kinematically forbidden.}
	\label{FigSemicond}
\end{figure}

\subsubsection{Calculation Methods}

The density functional theory calculations are performed using the Vienna $ab$ $initio$ simulation package.\cite{Kresse96a,Kresse96b} The pseudo-potentials are treated with the projector augmented wave (PAW) method\cite{Blochl94,KressePAW} with valence configurations of $6s^25p^65d^1$ for Dy and $5s^25p^4$ for Te. The plane-wave cut-off energy is set to 300 eV. All structures are optimized with the PBEsol functional\cite{PBEsol} with an energy convergence criteria of ($< 10^{-6}$ eV) where the Brillouin zone is sampled with Monkhorst-Pack meshes\cite{MP} of $12 \times 12 \times 6$ for pristine DyTe$_2$, $6 \times 6 \times 6$ for the \rfiveo~supercell structures, and
$6 \times 6 \times 3$ for the \rfived~supercell structures. The last Monkhorst-Pack grid is shifted by half grid point along $k_z$ direction to sample equivalent grid points. The Bloch states are then wannierized\cite{Mostofi,Marzari2012} with the modified Becke-Johnson (mBJ) potential\cite{mBJ1,mBJ2} in $\Gamma$-centered k-points meshes of $6 \times 6 \times 4$ for pristine DyTe$_2$ and $4 \times 4 \times 4$ for the \rfiveo~supercell structures. The effect of spin-orbit coupling is found to be negligible and not included here. The first Brillouin zone is sampled in $100 \times 100 \times 50$ grid points with the Wannier Hamiltonian and the tetrahedron method is employed to obtain the Fermi surface and nesting functions.

\subsubsection{Nesting function}

\begin{figure*}[h]
	\includegraphics[width=176mm]{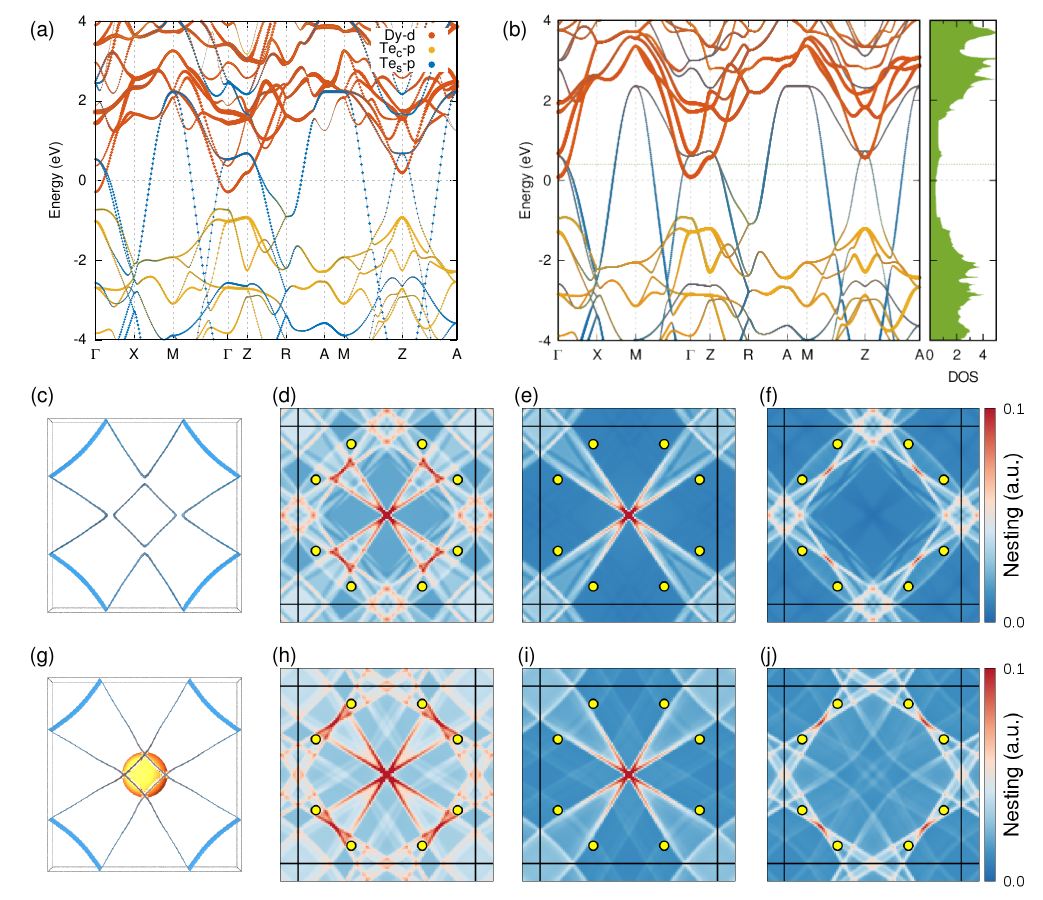}
	\caption{Electronic structure of pristine DyTe$_2$.	(a) Band structure along symmetry directions calculated with the PBEsol functional where the electron pocket derived from Dy-$d$ orbital appears at $\Gamma$. (b) Band structure calculated with Wannier Hamiltonian that is obtained with mBJ potential. The Dy-$d$ electron pocket at $\Gamma$ is lifted above the Fermi level. Right panel shows the density of states in units of (number of electrons per eV $\cdot$ 2Dy). Horizontal dashed line indicates electron doping level of $+0.2e$ per 2Dy. (c) Fermi surface and (d - f) three in-plane nesting	functions, $N^z_{0}$, $N^z_{P}$, and $N^z_{AP}$	at $E=0$ eV.	(g) Fermi surface and (h - j) three in-plane nesting functions, $N^z_{0}$, $N^z_{P}$, and $N^z_{AP}$ at $+0.2e$ doping level. The Te$_s$-$p$ and Dy-$d$ orbital characters in the Fermi surfaces in (c) and (g) are illustrated with blue and orange colors, respectively.}
	\label{FigNesting}
\end{figure*}

The Fermi surface nesting is often determined by the susceptibility $\chi(\mathbf {q})$.\cite{Shim_nesting} However, in this work we calculate the nesting function, $N(\mathbf{q})$, simply from the Fermi surface using the tetrahedron method, 
\begin{eqnarray*}
	N_{\gamma}(\mathbf{q}) &=& \sum_{i,j} {s_i s_j w_{\gamma}(\hat{\mathbf{v}}_i,\hat{\mathbf{v}}_j) \delta(\mathbf{k}_j - \mathbf{k}_i - \mathbf{q})},
\end{eqnarray*}
where the $i$ and $j$ indices run over the triangular facets of Fermi surface, $s_i$ is the area of $i$-th facet whose center is the $\mathbf{k}_i$ wavevector, and ${\gamma}$ denotes the type of weighting function $w_{\gamma}$, of the normalized group velocities $\hat{\mathbf{v}}_{i} = \mathbf{v}_{i} / |\mathbf{v}_{i}|$ of the $i$-th facets. We consider three types of weighting functions,
\begin{eqnarray*}
	w_{0}(\hat{\mathbf{v}}_i,\hat{\mathbf{v}}_j) &=& 1,\\
	w_{P}(\hat{\mathbf{v}}_i,\hat{\mathbf{v}}_j) &=& \frac{1}{2} (1+\hat{\mathbf{v}}_i \cdot \hat{\mathbf{v}}_j),\\
	w_{AP}(\hat{\mathbf{v}}_i,\hat{\mathbf{v}}_j) &=& \frac{1}{2} (1-\hat{\mathbf{v}}_i \cdot \hat{\mathbf{v}}_j),\\
\end{eqnarray*}
where mutual parallel and anti-parallel group velocities at facets $i$ and $j$ are distinguished with $w_{P}$ and $w_{AP}$ weighting functions, respectively. It is noteworthy that the parallel contribution is robust under shifts of the chemical potential whereas the anti-parallel contribution is relevant to the band gap opening. The in-plane nesting function is obtained by integrating along the $k_z$ direction as,
\begin{eqnarray*}
	N^z_{\gamma}(q_x,q_y) &=& \int_{0}^{2\pi} {dq_z N_{\gamma}(\mathbf{q})}.
\end{eqnarray*}
The $\mathbf{q}$ vector corresponding to the $\sqrt{5} \times \sqrt{5}$ modulation of DyTe$_2$ appears in the vicinity of mutual peaks of $N^z_{P}$ and $N^z_{AP}$ that explains the gap opening after the distortion.

\subsubsection{Formation energy}

\begin{figure*}[h]
	\includegraphics[width=176mm]{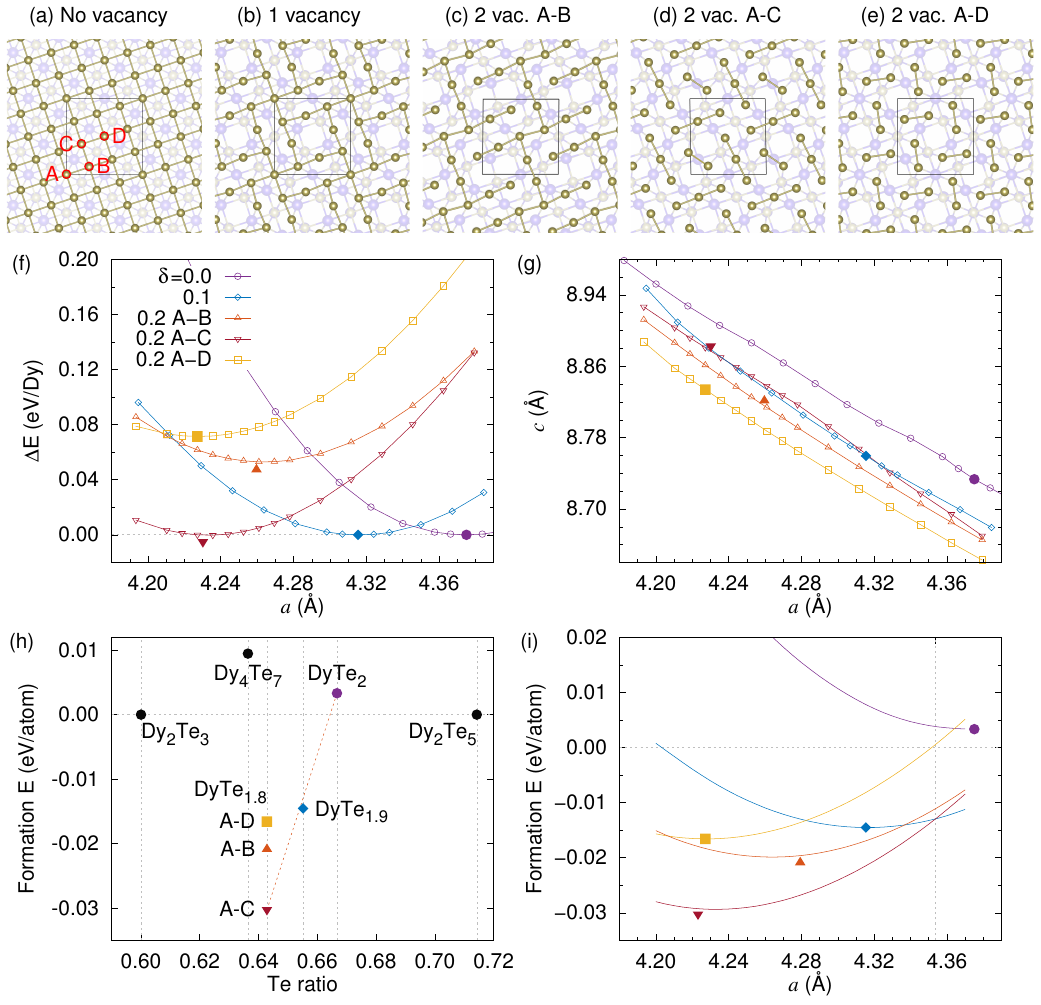}
	\caption{Calculated structural stability. (a - e) Te-bonding networks in the $\sqrt{5} \times \sqrt{5}$ supercell structures with various Te vacancy configurations. (f) Total energies of the various configurations shown in (a - e) as a function of epitaxial strain. All energies are relative to the lowest energy for each stoichiometry. (g) Out-of-plane, $c$ lattice constant under the epitaxial strain. (h) Calculated formation energy, referenced to Dy$_2$Te$_3$ and Dy$_2$Te$_5$ structures. (i) Formation energy as a function of epitaxial strain. Closed symbols in all panels denote the values obtained from fully relaxed structure without the epitaxial constraint ($a = b$). The two closed triangle symbols (A-B and A-C) breaks tetragonal symmetry ($a \neq b$) and are marked at $a' = \sqrt{ab}$.}
	\label{FigFormationE}
\end{figure*}

For the $\sqrt{5} \times \sqrt{5} \times 1$ supercell structures we consider Te mono- and di-vacancies per supercell, corresponding to $\delta=0.1$ and $0.2$, respectively. Without Te vacancies, the structural distortion is negligible indicating no phonon instability at the corresponding $\mathbf{q}$ vectors in the pristine DyTe$_2$. Introducing a single Te vacancy ($\delta=0.1$) causes rearrangement of the Te$_s$ ions and the formation of chain-like bonding networks. Two Te vacancies ($\delta=0.2$) can be arranged in three different configurations, referred to as A-B, A-C, and A-D shown in Fig.~\ref{FigFormationE}. The calculated relative total energies versus in-plane lattice constant for $\delta = 0.0$ (pristine), $0.1$ (mono-vacancy), and $0.2$ (di-vacancy) are shown in Fig.~\ref{FigFormationE} (f). The variation of lattice parameter $c$ versus in-plane lattice constant for the different Te-deficiency $\delta$ values are shown in Fig.~\ref{FigFormationE} (g). The closed symbols denote the ground state energy and the equilibrium lattice constants for all cases obtained by structural relaxation without epitaxial constraint. We find that the A-C di-vacancy configuration is the most stable. We note that A-B and A-C configurations break tetragonal symmetry and the total energies (closed triangle symbols) deviate from the epitaxial strain curves. In order to determine the structural stability between different stoichiometric phases, the formation energy is calculated relative to the Dy$_2$Te$_3$ and Dy$_2$Te$_5$ phases,
\begin{eqnarray*}
	E_{\textrm{Form}}^{(x,y)} = \frac{E_{\textrm{T}}^{(x,y)} - \frac{5x-2y}{16} E_{\textrm{T}}^{(8,12)} - \frac{-3x+2y}{8} E_{\textrm{T}}^{(4,10)}}{x+y}.
\end{eqnarray*}
Here, $E_\textrm{T}^{(x,y)}$ is the total energy of a structure consisting of $x$ Dy and $y$ Te atoms. The calculated formation energy [Fig.~\ref{FigFormationE} (h)] predicts that the $\delta=0.2$ A-C configuration is the most favored structure while DyTe$_2$ and Dy$_4$Te$_7$ phases are found to be less favorable states. Considering the distortion from pristine DyTe$_2$ to Te-deficient states, it is noteworthy that the formation energies of DyTe$_2$, DyTe$_{1.9}$, and DyTe$_{1.8, \textrm{A-C}}$ align on a line that indicates mixed occurrence between them.

It is thus an interesting question whether epitaxial strain which varies with film thickness can affect their formation energies due to the different equilibrium in-plane lattice constants, $a_{\delta=0.2} < a_{\delta=0.1} < a_{\delta=0.0}$. Figure~\ref{FigFormationE} (i) shows the formation energy as a function of epitaxial strain for the five structures that predicts a structural transition from $\delta=0.2$ A-C to $\delta=0.1$ phase with increasing epitaxial strain. The calculated threshold lattice constant is, however, larger than the maximum in-plane lattice constant in the experiment.

\begin{table}[h]
\label{Tab_c_mod}
\caption{Calculated total energies for the $\sqrt{5}\times\sqrt{5}\times2$ supercell structures with A-C divacancy configuration. One of the two Te square net layers is laterally shifted by $(d{_1}, d{_2}) = d{_1} \mathbf{a}^{'}_{1} + d{_2} \mathbf{a}^{'}_{2}$, where $\mathbf{a}^{'}_{i}$ are the in-plane translation lattice vectors of the supercell. For example, the vacancy sites A and C in Fig.~\ref{FigFormationE}(a) correspond to $(0.0, 0.0)$ and $(0.2, 0.4)$, respectively.}
\begin{tabular}{cc}
	Lateral shift & $\Delta E$ (meV/Dy) \\
	\hline
	(0.0, 0.0) &  0.0 \\
	(0.3, 0.1) & -6.0 \\
	(0.6, 0.2) & -8.4 \\
	(0.9, 0.3) & -5.8 \\
	(0.2, 0.4) & -7.3 \\
	(0.5, 0.5) & -10.8
\end{tabular}
\end{table}

\end{document}